\documentclass[conference]{IEEEtran}
\ifCLASSINFOpdf
\usepackage[pdftex]{graphicx}
\usepackage{array}
\graphicspath{{../pdf/}{../jpeg/}}
\DeclareGraphicsExtensions{.pdf,.jpeg,.png}
\else
\usepackage[dvips]{graphicx}
\graphicspath{{../eps/}}
\DeclareGraphicsExtensions{.eps}
\fi
\usepackage{algorithmic}
\usepackage{algorithm}

\begin{document}
%
\title{Flexible SQLf query based on fuzzy linguistic summaries}

\author{\IEEEauthorblockN{Ines Benali-Sougui$^1$, Minyar Sassi-Hidri$^2$, Amel Grissa-Touzi$^3$}
\IEEEauthorblockA{$^{1,2,3}$Universit\'e Tunis El Manar\\ $^{1,2}$Ecole Nationale d'Ing\'enieurs de Tunis\\Laboratoire Signal, Images et Technologies de l'information\\
BP. 37, Le Belv\`ed\`ere 1002, Tunis, Tunisia\\
$^{3}$Facult\'e des Sciences de Tunis\\Laboratoire d'Informatique, Programmation, Algorithmique et Heuristiques\\Campus Universitaire, Tunis 1060, Tunisia\\
$^1$ines.benali@gmail.com, $\{$$^2$minyar.sassi,$^3$amel.touzi$\}$@enit.rnu.tn}
}
\maketitle

\begin{abstract}
Data is often partially known, vague or ambiguous in many real world applications. To deal with such imprecise information, fuzziness is introduced in the classical model. SQLf is one of the practical language to deal with flexible fuzzy querying in Fuzzy DataBases (FDB). However, with a huge amount of fuzzy data, the necessity to work with synthetic views became a challenge for many DB community researchers. The present work deals with Flexible SQLf query based on fuzzy linguistic summaries. We use the fuzzy summaries produced by our Fuzzy-SaintEtiq approach. It provides a description of objects depending on the fuzzy linguistic labels specified as selection criteria.

Keywords-Fuzzy relational databases, Flexible querying, Data summarization, SQLf, Top k query, Fuzzy FCA.
\end{abstract}

\IEEEpeerreviewmaketitle

\section{Introduction}
In recent years, a lot of attention has been attracted to Fuzzy DataBases (FDB) that generalize the classical relational data model by allowing uncertain and imprecise information to be represented and manipulated. Data is often partially known, vague or ambiguous in many real world applications. Fuzziness is introduced in the classical model to deal with such imprecise information and several extensions of the model which are available in literature.

We are confronted more and more with the situation where applications need to manage fuzzy data and to profit their users from flexible querying \cite{Galindo,Ben Hssine,Bosc1998}.

In the last decades, a relational database language for fuzzy querying, called SQLf, has a big success for the description and the manipulation of the FDB (Fuzzy Data Bases) \cite{Bosc1995}. It extends the SQL by allowing the user to construct queries regarding atomic conditions defined by fuzzy sets. Each atomic condition combines satisfaction $\mu \in [0,1] $ to an attribute value.

In addition, SQlf limits the number of answers by using a quantitative calibration (the $k$ best responses or the top $k$ query) or qualitative calibration (the data that satisfy the query with an upper threshold $\alpha$).

However, the massive data reached today make necessary a better exploitation of the last. Several solutions have been proposed to solve this problem and to contribute in database summarization. Formal approaches are ones that have been proposed to surround this problem \cite{Bosc2000,Lee,Rashia}.

In \cite{BenAli-Sougui}, we have proposed a fuzzy linguistic summarization approach called Fuzzy-SaintEtiq using concept lattice which is the core of Formal Concept Analysis (FCA) \cite{Wille}. This approach is an extension of the SaintEtiQ model \cite{Mouaddib} to support the fuzzy data.It consists of two major steps: the first, called pre-processing step, considers a fuzzy clustering that permits the generation of fuzzy data partition associating the DB records or tuples to many clusters by means of memberships' degrees. This is a form of optimization as much in DB navigation as minimization of the domain expert risks compared to linguistic summarization proposed in \cite{Rashia}. The second step, called post processing, uses fuzzy concept lattice in order to generate conceptual hierarchy. Furthermore, querying summaries is crucial since it makes it possible to rapidly get a rough idea of the properties of tuples in a given relation.

Even that the interpretation of a summary is simple, it becomes difficult to predict a high number of summaries. So, we are faced to the problem of their use.

In this work, we propose to exploit the hierarchical summaries of fuzzy-SaintEtiq under flexible SQLf query. The theory of fuzzy sets \cite{Zadeh} used in the summarization process, allows flexible querying \cite{Voglozin2004}.

The rest of the paper is organized as follows: section 2 presents the basic concept of fuzzy FCA and the theoretical modeling of fuzzy queries. Section 3 presents an overview of our Fuzzy-SaintEtiq  approach presented in \cite{BenAli-Sougui}. Section 4 describes our new approach for the flexible SQLf query based on Fuzzy-SaintEtiq  approach. Section 5 concludes the paper and gives some future work.
\section{Preliminaries}
\subsection{Fuzzy FCA}
In this section, we discuss the Fuzzy FCA proposed by \cite{Thanh} which incorporates fuzzy logic into FCA to represent vague information.

\textbf{Definition 1}. A fuzzy formal context is a triple $K_{f} = (G,M, I = \varphi (G \times M))$ where $G$ is a set of objects, $M$ is a set of attributes, and $I$ is a fuzzy set on domain $G \times M$. Each relation $(g,m)\in I$ has a membership value $\mu(g,m)$ in $[0,1]$.

A fuzzy formal context can also be represented as a cross-table as shown in Table \ref{tab4}. The context has three objects representing three documents, namely $D_{1}$, $D_{2}$ and $D_{3}$. In addition, it also has three attributes, {\em Data Mining (D)}, {\em Clustering (C)} and {\em Fuzzy Logic (F)} representing three researchable topics. The relationship between an object and an attribute is represented by a membership value between 0 and 1. A confidence threshold $T$ can be set to eliminate relations that have low membership values \cite{Thanh}. Table \ref{tab4} shows the cross-table of the fuzzy formal context with $T = 0.5$.

\begin{table}[h!]
 \begin{center}
 \small{
   \tabcolsep = 0.5\tabcolsep
\caption{Fuzzy Formal Context with $T=0.5$} \label{tab4}
   \begin{tabular}{|l|c|c|c|}
   \hline
   & D & C & F \\
         \hline
  D1 & 0.8 & - & 0.61 \\
  D2 & 0.9 & 0.85 & - \\
  D3 & - & - & 0.87 \\
     \hline
        \end{tabular}
        }
 \end{center}
\end{table}

Each relationship between the object and the attribute is represented as a membership value in fuzzy formal context, then the intersection of these membership values should be the minimum of them, according to fuzzy theory \cite{Zadeh}.

\textbf{Definition 2}. Fuzzy formal concept: Given a fuzzy formal context $K_{f} = (G,M, I = \varphi (G \times M))$ and a confidence threshold $T$, we define $A\ast= \{m\in M|\forall g\in A: \mu(g, m) \geq T\}$ for $A \subseteq G$ and $B\ast = \{g\in G|\forall m\in B: \mu(g,m)\geq T\} $ for $B \subseteq M$.

A fuzzy formal concept (or fuzzy concept) of a fuzzy formal context $K_{f} = (G,M, I = \varphi (G \times M))$ with a confidence threshold $T$ is a pair $(A_{f} = \varphi(A), B)$ where $A \subseteq G$, $B \subseteq  M$, $A \ast = B$ and $B\ast = A$. Each object $g=\varphi(A)$ has a membership $\mu_{g}$ defined as $\mu_{g} = min(\mu(g,m))$ and $m\in B$ where $\mu(g,m)$ is the membership value between an object $g$ and an attribute $m$, which is defined in $I$. Note that if $B= \{\}$ then $\mu_{g} = 1$ for every $g$.

\textbf{Definition 3}. Let $(A_{1}, B_{1})$ and $(A_{2}, B_{2})$ be two fuzzy concepts of a fuzzy formal context $(G, M, I)$. $( \varphi(A_{1}), B_{1})$ is the sub-concept of $(\varphi (A_{2}), B_{2})$, denoted as $(\varphi(A_{1}), B_{1}) \leq (\varphi (A_{2}), B_{2})$, if and only if $\varphi(A_{1}) \subseteq  \varphi(A_{2})$ ($\Longleftrightarrow B_{2} \subseteq B_{1}$). Equivalently, $(A_{2}, B_{2})$ is the super-concept of $(A_{1}, B_{1})$.

\textbf{Definition 4}. A fuzzy concept lattice of a fuzzy formal context $K$ with a confidence  threshold $T$ is a set $F (K)$ of all fuzzy concepts of $K$ with the partial order $\leq$ with the confidence threshold $T$.

\textbf{Definition 5}. The similarity of a fuzzy formal concept $K_{1}= (\varphi (A_{1}), B_{1})$ and its sub-concept $K_{2}= ( \varphi(A_{2}), B_{2})$ is defined as:

\begin{equation}
\small{
E(K_{1},K_{2}) = \frac{|\varphi (A_{1}) \cap \varphi (A_{2})|}{| \varphi (A_{1})  \cup \varphi(A_{2})|}
}
\end{equation}

\begin{figure}[!ht]
\centering
\includegraphics[scale=0.35]{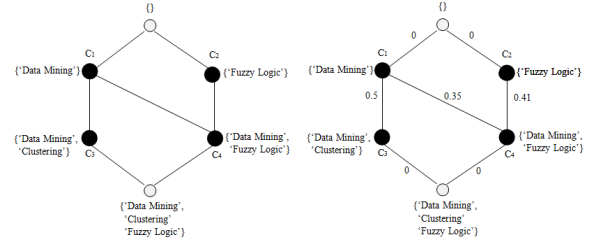}
\caption{(a) A concept lattice generated from traditional FCA. (b) A fuzzy concept lattice generated from Fuzzy FCA}
\label{fig1}
\end{figure}

Figure \ref{fig1} (a) gives the traditional concept lattice generated from table \ref{tab4}, in which crisp values {\em Yes} and {\em No} are used instead of membership values. Figure \ref{fig1} (b) gives the fuzzy concept lattice generated from the fuzzy formal context given in Table \ref{tab4}.
\subsection{Fuzzy queries' modeling}
The SQLf \cite{Bosc1995} language extends the SQL language by allowing the user to construct queries on atomic conditions defined by fuzzy sets. Each atomic condition combines a satisfaction level $  \mu \in [0, 1]$ with an attribute value. For all the attributes of an n-uplet, the semantics of degrees are the same which involve that all criteria are commensurable. SQLf query has the following syntax:\\
{\em \textbf{Select} [distinct] [$k$ $\mid$ $\alpha$ $\mid$k,$\alpha$ ] $\langle$ attribut $\rangle$}\\
{\em \textbf{From} $\langle$ strict relation $\rangle$}\\
{\em \textbf{Where} $\langle$ fuzzy condition $\rangle$}

where $\langle$ fuzzy condition $\rangle$ can incorporate blocks of queries nested or partitioned. Parameters $k$ and $\alpha$ from Select clause limit the number of answers by using a quantitative calibration($k$ best responses) or qualitative calibration(the data that satisfy the query with a threshold greater than $\alpha$).

\textbf{Example}. Let be consider the employee DB relation presented in table \ref{tab5} and the following query: {\em Finding employees about 40 years and having a high income}.

\begin{table}[h!]
 \begin{center}
   \tabcolsep = 1\tabcolsep
\caption{Relationship Employee} \label{tab5}
   \begin{tabular}{|l|c|c|c|}
   \hline
 Id & Name & Age & Income \\
         \hline
  1 & Pierre & 38 & 2900 \\
  2 & Jean & 37 & 2800 \\
  3 & Yvette & 42 &2700 \\
     \hline
   \end{tabular}
 \end{center}
\end{table}
About 40 and high income are the selection criteria defined by fuzzy sets. In particular, for the n-uplets of the relation employees we have used {\em High income={0.9/2900, 0.8/2800, 0.7/2700}}, {\em About 40={0.8/38, 0.6/37, 0.8/42}}.

Each n-uplet is associated with a vector that represents its position regarding the atomic conditions. Thus, the final result is evaluated and calculated using a standard triangular (eg min) to express a conjunction between the criteria. We then obtain: $ {Pierre(0.8)> Yvette(0.7)> Jean (0.6)}$  means that the tuple 1 is preferred to tuple 3 which is preferred to tuple 2.

With SQLf language, preferences are considered only as constraints and are taken into account through the expression of fuzzy predicates commensurable, modeled using fuzzy sets of values more or less satisfactory.

Consequently, the results are totally ordered. Such predicates can be combined using a rich platform operators of logical fuzzy, with some reflecting the effects of such compensation or relative importance between criteria, have no counterpart in Boolean logic. Unlike other approaches (such as Preference SQL), data selection and order preferences operation are processed simultaneously.

Besides, {\em top-k query} has attracted much interest in many different areas such as network and system monitoring \cite{Babcock}, information retrieval \cite{Kimelfeld}, sensor networks\cite{Silberstein,Wu}, large databases \cite{GTouzi}, multimedia databases \cite{Chaudhuri}, spatial data analysis \cite{Ciaccia,Hjaltason}, P2P systems \cite{Akbarinia}, data stream management systems \cite{Moutatidis} etc.

The main reason for such interest is that they avoid overwhelming the user with large numbers of uninteresting answers which are resource-consuming.

The problem of answering top $k$ queries can be modeled as follows \cite{Fagin}. Suppose that we have m lists of n data items such that each data item has a local score in each list and the lists are sorted according to the local scores of their data items. And each data item has an overall score which is computed based on its local scores in all lists using a given scoring function. Then the problem is to find the $k$ data items whose overall scores are the highest.
\section{Fuzzy data summarization}
\subsection{Overview of Fuzzy-SaintEtiq approach}
\begin{figure}[!ht]
\centering
\includegraphics[scale=0.40]{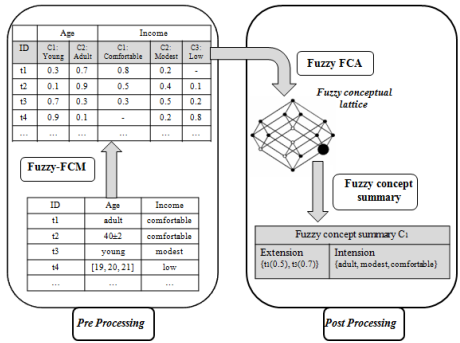}
\caption{The Overall process of Fuzzy-SaintEtiq}
\label{fig2}
\end{figure}

In \cite{BenAli-Sougui}, we have proposed a fuzzy linguistic summarization approach Fuzzy-SaintEtiq which is based on FCA-based Summary model \cite{Sassi}. It takes the database records and provides knowledge. Figure \ref{fig2}  gives the system architecture.
\\
The summarization act considered like a process of knowledge discovery from database, in the sense that it is organized according to two following principal steps. The preprocessing step which organizes the FDB records in homogeneous clusters having common properties. This step gives a certain number of clusters for each attribute. Each tuple has values in the interval [0..1] representing these membership degrees according to the formed clusters. Linguistic labels, which are fuzzy partitions, will be assigned on attribute's domain.
For the classification on these fuzzy data, a new algorithm, called Fuzzy-FCM \cite{GrissaTouzi} has been proposed. It's an extension of FCM algorithm in order to support different types of data represented by GEFRED model \cite{Galindo}. The second step, called the post processing, takes into account the result of the fuzzy clustering on each attribute, visualizes by using the fuzzy concepts lattices. Then, it nests them in a fuzzy nested lattice. Finally, it generalizes them in a fuzzy lattice associating all records in a simple and hierarchical structure. Each lattice node is a fuzzy concept which represents a concept summary. This structure defines summaries at various hierarchical levels.

For a more formal expression of a query, let be consider:
\begin{itemize}
\item $A=\{A_{1},A_{2},...,A_{k}\}$  is a set of attributes,
\item$A_{k}$ is the $k^{th}$ attribute which appears is query $Q$,
\item $R(A)$ is the relation whose tuples are summarized,
\item $t_{i}$ is the $i^{th}$ tuple, $i\in 1...N$,
\item $L_{k}=\{l_{1k},...,l_{jk}\}$ is a set of linguistic terms of attribute $A_{k}$,
\item $\mu_{ijk}$ is a membership degree of  tuple $t_{i}$ (record) to the linguistic term $l_{jk}$ of attribute  $A_{k}$,
\item $R_{Z_{f}}$ is the sub set of involved tuples in summary $z_{m}$,
\item $I_{Z_{f}}$ is the sub set of linguistic terms which appears in summary $z_{f}$,
\item $Z_{f}=(R_{z_{f}},I_{Z_{f}})$ is the concept summary, $R_{zm}$ and $I_{Z_{f}}$ are respectively the intension and the extension of the concept,
\item $level$ is the level of the concept summary in the concept lattice,
\item $|R_{Z_{f}}|$ is the number of candidates tuples in $R_{Z_{f}}$.
\end{itemize}
\subsection{Illustrative example}
Let consider a  relation  \textbf{$R=$($IdTuple$, $Age$, $Income$, $ProfessionalBackground$)} from an FDB Employees table. Table \ref{tab1} gives a sample of FDB employees table.

\begin{table}[h!]
 \begin{center}
   \tabcolsep = 0.5\tabcolsep
\caption{Data sample} \label{tab1}
   \begin{tabular}{|l|c|c|c|}
   \hline
 IdTuple &  Age& Income& ProfessionalBackground    \\
         \hline
$t_{1}$ & [38,39,40]&	950&	10 \\
$t_{2}$ & Adult	&650&	5  \\
$t_{3}$ & Young	&$700\pm20$&	3  \\
$t_{4}$ & Adult	&Poor&	20 \\
$t_{5}$ & 38&	Poor	&7  \\
$t_{6}$ &$40\pm2$	&Comfortable	&12   \\
     \hline
   \end{tabular}
 \end{center}
\end{table}

Each cluster of a partition is labeled by linguistic descriptor provided by a domain expert. For example, the fuzzy label {\em Young} belongs to a partition built on the domain of age attribute. Linguistic variables associated with the attributes of $R$. These linguistic variables constitute the new attribute domains used for the rewriting of tuples in the summarization process. For clusters generation, we carry out a fuzzy clustering \cite{vanistri:inter} while benefiting from fuzzy logic. This operation makes it possible to generate, for each attribute, a set of membership degrees. In fact, several fuzzy clustering algorithms have been proposed in the literature \cite{sassi:afss,pradeep:ij}.

Table \ref{tab2} presents the results of fuzzy clustering applied to {\em Age} and {\em Income} attributes. For {\em Age} attribute, fuzzy clustering generates two clusters whereas, for Income attribute, there are three clusters. The minimal value (resp. maximal) of each cluster corresponds to the lower (resp. higher) interval terminal of the values of this last. Each cluster of a partition is labeled with a linguistic descriptor provided by the user or a domain expert. For this, the following abbreviations are used:
\begin{itemize}
\item For {\em Age} attribute: {\em YA} (Young Age) and {\em AA} (Adult Age).
\item For Income attribute: {\em PI} (Poor Income), {\em MI} (Modest Income) and {\em CI} (Comfortable Income).
\item For Professional background attribute: {\em A} (Associate), {\em E} (Expert) and {\em S} (Senior).
\end{itemize}

Table \ref{tab2} gives the result of fuzzy clustering from data in table \ref{tab1}.

\begin{table}[h!]
 \begin{center}
 \small{
   \tabcolsep = 0.5\tabcolsep
\caption{Data clustering} \label{tab2}
   \begin{tabular}{|l|c|c|c|}
   \hline
 IdTuple &  Age& Income& ProfessionalBackground    \\
         \hline

$t_{1}$ & $YA^{0.5}$, $AA^{0.5}$&	$MI^{0.6}$, $CI^{0.4}$&	$A^{0.7}$, $E^{0.3}$ \\
$t_{2}$ & $YA^{0.4}$, $AA^{0.6}$&	$PI^{0.4}$, $MI^{0.6}$	&$A^{0.5}$, $E^{0.5}$  \\
$t_{3}$ & $YA^{0.7}$, $AA^{0.3}$	&$MI^{0.8}$	&$A^{0.8} $ \\
$t_{4}$ & $YA^{0.2}$, $AA^{0.8}$&	$PI^{0.6}$, $MI^{0.4}$	&$E^{0.3}$, $S^{0.7}$\\
$t_{5}$ & $YA^{0.6}$, $AA^{0.4}$	&$PI^{0.7}$	&$A^{0.7}$, $E^{0.3}$  \\
$t_{6}$ & $YA^{0.5}$, $AA^{0.5}$&	$CI^{0.8}$	&$E^{0.4}$, $S^{0.6}$  \\
     \hline
   \end{tabular}
   }
 \end{center}
\end{table}
\section{An SQLf-based flexible summary querying}
 A flexible querying process of a DB can be divided into three steps \cite{Larsen}: extension of criteria, selection of results and ordering. The first step uses similarities between values to extend the criteria. It allows graded semantics for any criterion, which can now express {\em around 20} instead of being limited to the binary semantics of {\em equal to 20} or {\em between 18 and 22}. The second step, namely the selection of results, determines which data will participate in the answer to the query. The last step (ordering) follows the extension of criteria. It discriminates the results on the basis of their relative satisfaction to the graded semantics: a value of 20 is better ranked than a value of 18.

The following works, which are the research of flexible query in FDB, exemplify the use of fuzzy sets. They are essentially characterized by a tuple-oriented processing, the possibility to define new terms and especially the use of satisfaction degrees to extract the top-k query.

\subsection{Fuzzy query expression}
In the query SQLf we have two parameter $\alpha$ and $k$. As previously said parameters $k$ and $\alpha$ from Select clause limits the number of answers by using a quantitative calibration($k$ best responses) or qualitative calibration(the data that satisfy the query with a threshold greater than $\alpha$) .

The parameter k is given by the user and the parameter $\alpha$ is calculated depending on the number of clusters which involved in condition of Select query.

\begin{equation}
\small{
\alpha=\frac{1}{max(NClus)}
}
\end{equation}

where $NClus$ is the number of clusters involved in the condition of select clause.

Let us consider the example in the table \ref{tab1}. The queries are as follows:\\
{\em \textbf{Q$_{1}$}}\\
{\em \textbf{Select} 3 0.5 Income, ProfessionalBackground}\\
{\em \textbf{From} Employees \textbf{Where} Age IN (Young);}

and\\
{\em \textbf{Q$_{2}$}}\\
{\em \textbf{Select} 3 0.3 ProfessionalBackground}\\
{\em \textbf{From} Employees}\\
{\em \textbf{Where} Income IN (Comfortable) AND Age IN (Young);}

In a query, descriptors like Young, Comfortable in $Q_{1}$ and $Q_{2}$ are called {\em required characteristics} and embody the properties that a record must consider them as an element of the answers. A query also defines the attributes for which required characteristics exist. The set of these input attributes is denoted by $Inputs(A_{Q})$. The expected answer is a description over a set of other attributes, denoted by $Outputs(A_{Q})$. It is the complement of $Inputs(A_{Q})$ relatively to $A_{Q}$ (the set of attributes appears in the query $Q$):
\begin{equation}
\small{
Inputs(A_{Q})\cup Outputs(A_{Q})=A
}
\end{equation}

and
\begin{equation}
\small{
Inputs(A_{Q})\cap Outputs(A_{Q})=\emptyset
}
\end{equation}

Hence a query $Q$ defines not only a set $Inputs(A_{Q})$ of input attributes  but also for each attribute $A_{k}$, the set $L_{A_{k}}(Q)$ of its required characteristics which define the set of linguistic terms of attribute $A_{k}$ appears query $Q$. The set of sets $L_{A_{k}}(Q)$ is denoted by $L(Q)$.

For example, for $Q_{2}$, this set is determined as follows:
\begin{itemize}
\item $Inputs(A_{Q_{2}})=\{Income,Age\}$;
\item $Outputs(A_{Q_{2}})=\{ProfessionalBackground\}$;
\item $L_{Income}(Q_{2})=\{RC\}$, $L_{Age}(Q_{2})=\{AJ\}$;
\item $L(Q_{2})=\{L_{Income}(Q_{2},L_{Age}(Q_{2})\}$.
\item The degree of membership to this query is 0.3 and the number of the desired result is 5.
\end{itemize}
\subsection{Fuzzy query rewriting}
The query rewriting in a logical proposition $P_{f}(Z_{f}, Q$) used to qualify the link between the fuzzy summary $Z_{f}$ and the query $Q$. $P_{f}(Z_{f},Q)$ is in a conjunctive form in which all descriptors are literals. Then, each set of descriptors yields one corresponding clause.Thereafter we will apply an $\alpha$-cut on this new form with the parameters $\alpha$ is calculated previously.
\\
This form is defined as follows:
\begin{flushleft}
\small{
$min(l_{11}\vee l_{21}\vee ...\vee l_{j1})(x), min(l_{12}\vee l_{22}\vee...\vee l_{j2})(x),...,$

$min(l_{1k}\vee l_{2k} \vee...\vee l_{jk})(x)\geq\alpha$
}
\end{flushleft}
\begin{flushleft}
\small{
$\Longleftrightarrow$ $(l_{11}\vee l_{21} \vee ... \vee l_{j1}) (x)\geq\alpha$,

$(l_{12} \vee l_{22} \vee ... \vee l_{j2})(x)\geq\alpha$
 ,... ,$(l_{1k} \vee l_{2k} \vee ... \vee l_{jk})(x)\geq\alpha$
}
\end{flushleft}
\begin{flushleft}
\small{
$\Longleftrightarrow$ $l_{11}(x)\geq\alpha  $
or $l_{21}(x)\geq\alpha$ or $l_{j1}(x)\geq\alpha$, $l_{12}(x)\geq\alpha$ or $l_{22}(x)\geq\alpha$ or $l_{j2})(x)\geq\alpha$, $l_{1k}(x)\geq\alpha$ or $l_{2k}(x)\geq\alpha$ or $l_{jk}(x)\geq\alpha$
}
\end{flushleft}
\begin{flushleft}
\small{
$\Longrightarrow$
$P(Q)=(\alpha-cut(l_{11})\vee\alpha-cut(l_{21}) \vee...\vee \alpha-cut(l_{j1}) )$

$\wedge(\alpha-cut(l_{12}) \vee \alpha-cut(l_{22}) \vee...\vee\alpha-cut(l_{j2}))\wedge...\wedge$

$(\alpha-cut(l_{1k}) \vee \alpha-cut( l_{2k}) \vee...\vee \alpha-cut(l_{jk}))$
}
\end{flushleft}
\textbf{Example:} Let be consider the query $Q_{3}$:\\
{\em \textbf{Select} 3 0.3  ProfessionalBackground}\\
{\em \textbf{From} Employees}\\
{\em \textbf{Where} Age IN (Young, Adult)}\\
{\em \textbf{And} Income IN (Poor, Modest);}

We have then:
\begin{itemize}
\item $Inputs(A_{Q_{1}})=\{Age,Income\}$;
\item $L_{Age}=\{YA,AA\}$;
\item $L_{Income}=\{PI,MI\}$;
\item The degree of membership to this query is 0.3 and the number of the desired result is 3;
\item $P(Q_{3})=(0.3-cut(YA) \vee 0.3-cut(AA))$ $\wedge (0.3-cut(PI)$ $\vee 0.3-cut(MI))$.
\end{itemize}

Let be consider $v_{f}$ the valuation function. It is obvious that the valuation of $P_{f}(Q)$ depends on the summary $Z_{f}$. Thus $v_{f}(P_{f}(Q)_{Z_{f}})$ denotes the valuation of $P_{f}(Q)$ in the context of $Z_{f}$. $L_{A_{i}}(Z_{f})$  the set of linguistic terms that appear in $Z_{f}$. We can distinguish between three assumptions:

\begin{itemize}
\item $Coresp(Z_{f}, Q)=Exact$ : $ v_{f}(P_{f}(Q)_{Z_{f}})=true$ and $L_{A_{i}}(Z_{f})\subseteq L(Q)$ : All tuples including in  $Z_{f}$ verify the query $Q$;
\item $Coresp(Z_{f}, Q)=False$ : $v_{f}(P_{f}(Q)_{Z_{f}})=false$ : $L_{A_{i}}(Z_{f})\neq L(Q)$ : Linguistic terms appear in $Z_{f}$ do not correspond to terms in query $Q$;
\item $Coresp(Z_{f}, Q)=Indecision$ : $\exists i$, $L_{A_{i}}(Z_{f})- L(Q)\neq \emptyset$: There are some tuples in  $Z_{f}$ satisfying $Q$.
\end{itemize}
These situations reflect a global view of the matching of a fuzzy summary $Z_{f}$ with a query $Q$.
\subsection{Fuzzy k-query}
\begin{figure}[!ht]
\begin{center}
\includegraphics[scale=0.43]{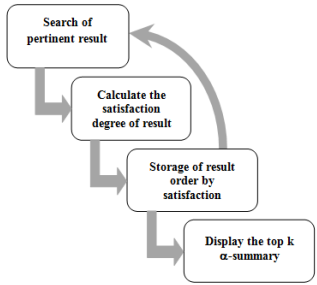}
\caption{An SQLf-based flexible summary querying approach step} \label{fig3}
\end{center}
\end{figure}

The idea of the Fuzzy k-query algorithm is to search using the summary concept; which summary responds to the query and calculate their satisfaction degree. Then we will insert them in a list order by satisfaction degree. We will repeat these steps until ensure that there is not a branch in the summary concept that satisfies the query. Finally we can display the top $k$ $\alpha-summary$ from the list of results.
Figure \ref{fig3} shows the principle of our approach.
\\
\textbf{Definition 6.} An $\alpha-summary$, denoted as $\alpha-Z_{f}$, is a fuzzy summary $Z_{f}=(R_{Z_{f}}, I_{Z_{f}})$ in which $R_{Z_{f}}$ is a collection of candidate records $R_{Z_{f}}=\{t_{1}, t_{2},..., t_{N}\}$ which represents the extent and $I_{Z_{f}}$ is the intent. Each tuple $t_{i}$ of $R_{Z_{f}}$ existing in the $\alpha-summary$ has a membership value $\mu(t_{i})\geq\alpha$. It can be formulated as follows:
\begin{equation}
\small{
\alpha-Z_{f}=\{\forall~t_{i} \in Z_{f}| \mu (t_{i})\geq \alpha\},~with~~\mu(t_{i})\in[0,1]
}
\end{equation}
\subsection{Query evaluation}
In this section, we try to evaluate the proposed approach. For this, the searching procedure should take into account all fuzzy summaries in the concept lattice that correspond to the query $Q$. The evaluation procedure is based on a generalization search and relies on the property of the concept lattice hierarchy.
The algorithme \ref{alg1} describes the different steps of the {\em fuzzy k-query} function where $k$ is the number of answers, $Z_{result}$ is a list of all $\alpha-summary$ responding to a query $Q$.
\begin{algorithm}
\caption{Fuzzy k-query($Z_{f}$, $k$, $\alpha$, $Q$)}
\label{alg1}
\small{
\begin{algorithmic}[1]
\REQUIRE $Z_{f}$ the fuzzy concept summary, $k$ the number of answers, $\alpha$ the threshold greater data to satisfy the query $Q$.
\ENSURE Result list of the top $k$ $\alpha-summary$ responding to the query $Q$.
\STATE $Result \Leftarrow \emptyset$ List of summary with their satisfaction degrees
\STATE ${Z_{result} \Leftarrow PertinentResult(Z, 0, \alpha, Q )}$
\STATE ${Z_{result}.first()}$
\FOR {$i = 1 \to k$}
\STATE ${Result.info() \Leftarrow Z_{result}.info()}$
\STATE ${Z_{result}.next()}$
\STATE ${Result.next()}$
\ENDFOR
\end{algorithmic}
}
\end{algorithm}

The algorithm \ref{alg2} describes the different steps of the {\em PertinentResult} procedure.
\begin{algorithm}[ht!]
\caption{PertinentResult($Z_{f}$, $level$,  $\alpha$, $Q$)}
\label{alg2}
\small{
\begin{algorithmic}[1]
\REQUIRE $Z_{f}$  the fuzzy concept summary, $level$ the current level of summary, $\alpha$ the threshold greater data to satisfy the query $Q$.
\ENSURE $Z_{result}$ list of all $\alpha-summary$ responding to the query $Q$.
\STATE $Z_{result} \Leftarrow \emptyset$
\IF {$Coresp(Z, Q) = Exact$}
\STATE ${x.degree \Leftarrow Calcul-SD(Z, level )}$
\STATE ${x.sum \Leftarrow \alpha-summary(Z,\alpha)}$
\IF {${Z_{result}.Empty()}$}
\STATE ${Z_{result}.Insertprevious(x)}$
\ELSE
\STATE ${Z_{result}.first()}$
\ENDIF
\WHILE {$not(Z_{result}.offlist())$}
\IF {${Z_{result}.degree < x.degree}$}
\STATE ${Z_{result}.Insertprevious(x)}$
\ENDIF
\STATE ${Z_{result}.next()}$
\ENDWHILE
\ELSE
\IF {$Coresp(Z, Q) = indecision$)}
\FORALL {Fuzzy summary $z_{f}$ of $Z_{f}$}
\STATE $ {Level \Leftarrow level+1}$
\STATE $Z_{result}\Leftarrow Z_{result}+PertinentResult(Z_{f},level, \alpha,Q)$
\ENDFOR
\ENDIF
\ENDIF
\end{algorithmic}
}
\end{algorithm}
\\
$Calcul-SD $ is the function to calculate the satisfaction degree $SD$ of fuzzy summary $Z_{f}$.This degree is the $max$ of the road that lets us find $Z_{f}$. It is determined as follows:
\begin{equation}
\small{
SD= max (\sum Fuzzy-score(Z_{j}, Z_{j+1}))
}
\end{equation}with $j=1..p$, p the current level of $Z_{f}$ and
 \begin{equation}
 \small{
Fuzzy-score(Z_{p}, Z_{p+1})=\frac{| (Z_{p})  \cap  (Z_{p+1})|}{| ( Z_{p})  \cup  ( Z_{p+1})|}
}
\end{equation}
$Coresp(Z, Q)$ is the function that allows to test the correspondence between the summary $Z_{f}$ and the query $Q$ that we have seen previously.

$\alpha-summary(Z,\alpha )$ is the function  that uses the definition of $\alpha-summary$; the result is the $ \alpha-Z=\{ \forall t\in Z,  \mu(t)\geq \alpha \}$.
\\

The result of applying the algorithm on the concept lattice for queries $Q_{1}$, $Q_{2}$ and $Q_{3}$ is given in table \ref{tab3}.

\begin{table}[ht!]
 \begin{center}
 \small{
   \tabcolsep = 0.5\tabcolsep
\caption{Top $k$ $\alpha-summary$} \label{tab3}
   \begin{tabular}{|l|p{7cm}|}
   \hline
 Query &       $\alpha-Summary$     \\
         \hline \hline

$Q_{1}$ &$\alpha-z_{13}$=\{$t_{1}^{0.5}$, $t_{3}^{0.7}$, $t_{5}^{0.6}$, $t_{6}^{0.5}$ \} \\
$Q_{2}$ & $\alpha-z_{42}=\{t_{1}^{0.4}\}$,
  $\alpha-z_{34}$=\{$t_{1}^{0.4}$, $t_{6}^{0.8}$\}   \\
$Q_{3}$ & $\alpha-z_{21}$=\{$t_{1}^{0.5}$, $t_{2}^{0.6}$, $t_{4}^{0.4}$\},
  $\alpha-z_{22}$=\{$t_{1}^{0.5}$, $t_{2}^{0.4}$, $t_{3}^{0.7}$ \},
  $\alpha-z_{23}$=\{$t_{2}^{0.4}$, $t_{4}^{0.6}$, $t_{5}^{0.7}$\}\\
        \hline
   \end{tabular}
   }
 \end{center}
\end{table}
\section{Conclusion}
In this paper we have presented a flexible SQLf query based on fuzzy linguistic summaries. Because, the SQLf query limits the number of answers by using a quantitative calibration or qualitative calibration, the mechanism of  Fuzzy k-query explores a hierarchy of fuzzy summaries. Each fuzzy summary, represented by a fuzzy concept, performs a comparison with set-based query descriptors from a predefined vocabulary and applies an $\alpha-cut$. The result of this comparison determines whether the abstract will be part of the answer but it also conditions the exploration of a part of the hierarchy. Then using the satisfaction degree that will be calculated for each result allows us to return the {\em top k result}.

In brief, our objective is to found a result even in the case of the absence of summaries corresponding
strictly to the query. To accommodate this target we will introduce a new kind of repaired query. Reparation is based on the idea that there could be a result semantically close to the query. To find these results, the query is modified using the best fuzzy summaries which is the near value answering the query.

\end{document}